\documentstyle[11pt,psfig,twoside]{article}

\begin{document}

\newcommand{\dd}{\,{\rm d}}
\newcommand{\ie}{{\it i.e.},\,}
\newcommand{\etal}{{\it et al.\ }}
\newcommand{\eg}{{\it e.g.},\,}
\newcommand{\cf}{{\it cf.\ }}
\newcommand{\vs}{{\it vs.\ }}
\newcommand{\zdot}{\makebox[0pt][l]{.}}
\newcommand{\up}[1]{\ifmmode^{\rm #1}\else$^{\rm #1}$\fi}
\newcommand{\dn}[1]{\ifmmode_{\rm #1}\else$_{\rm #1}$\fi}
\newcommand{\upd}{\up{d}}
\newcommand{\uph}{\up{h}}
\newcommand{\upm}{\up{m}}
\newcommand{\ups}{\up{s}}
\newcommand{\arcd}{\ifmmode^{\circ}\else$^{\circ}$\fi}
\newcommand{\arcm}{\ifmmode{'}\else$'$\fi}
\newcommand{\arcs}{\ifmmode{''}\else$''$\fi}
\newcommand{\MS}{{\rm M}\ifmmode_{\odot}\else$_{\odot}$\fi}
\newcommand{\RS}{{\rm R}\ifmmode_{\odot}\else$_{\odot}$\fi}
\newcommand{\LS}{{\rm L}\ifmmode_{\odot}\else$_{\odot}$\fi}

\newcommand{\Abstract}[2]{{\footnotesize\begin{center}ABSTRACT\end{center}
\vspace{1mm}\par#1\par
\noindent
{~}{\it #2}}}

\newcommand{\TabCap}[2]{\begin{center}\parbox[t]{#1}{\begin{center}
  \small {\spaceskip 2pt plus 1pt minus 1pt T a b l e}
  \refstepcounter{table}\thetable \\[2mm]
  \footnotesize #2 \end{center}}\end{center}}

\newcommand{\TableSep}[2]{\begin{table}[p]\vspace{#1}
\TabCap{#2}\end{table}}

\newcommand{\FigCap}[1]{\footnotesize\par\noindent Fig.\  %
  \refstepcounter{figure}\thefigure. #1\par}

\newcommand{\TableFont}{\footnotesize}
\newcommand{\TableFontIt}{\ttit}
\newcommand{\SetTableFont}[1]{\renewcommand{\TableFont}{#1}}

\newcommand{\MakeTable}[4]{\begin{table}[htb]\TabCap{#2}{#3}
  \begin{center} \TableFont \begin{tabular}{#1} #4 
  \end{tabular}\end{center}\end{table}}

\newcommand{\MakeTableSep}[4]{\begin{table}[p]\TabCap{#2}{#3}
  \begin{center} \TableFont \begin{tabular}{#1} #4 
  \end{tabular}\end{center}\end{table}}

\newenvironment{references}%
{
\footnotesize \frenchspacing
\renewcommand{\thesection}{}
\renewcommand{\in}{{\rm in }}
\renewcommand{\AA}{Astron.\ Astrophys.}
\newcommand{\AAS}{Astron.~Astrophys.~Suppl.~Ser.}
\newcommand{\ApJ}{Astrophys.\ J.}
\newcommand{\ApJS}{Astrophys.\ J.~Suppl.~Ser.}
\newcommand{\ApJL}{Astrophys.\ J.~Letters}
\newcommand{\AJ}{Astron.\ J.}
\newcommand{\IBVS}{IBVS}
\newcommand{\PASP}{P.A.S.P.}
\newcommand{\Acta}{Acta Astron.}
\newcommand{\MNRAS}{MNRAS}
\renewcommand{\and}{{\rm and }}
\section{{\rm REFERENCES}}
\sloppy \hyphenpenalty10000
\begin{list}{}{\leftmargin1cm\listparindent-1cm
\itemindent\listparindent\parsep0pt\itemsep0pt}}%
{\end{list}\vspace{2mm}}

\def\TYLDA{~}
\newlength{\DW}
\settowidth{\DW}{0}
\newcommand{\dw}{\hspace{\DW}}

\newcommand{\refitem}[5]{\item[]{#1} #2%
\def\REFARG{#3}\ifx\REFARG\TYLDA\else, {\it#3}\fi
\def\REFARG{#4}\ifx\REFARG\TYLDA\else, {\bf#4}\fi
\def\REFARG{#5}\ifx\REFARG\TYLDA\else, {#5}\fi.}

\newcommand{\Section}[1]{\section{#1}}
\newcommand{\Subsection}[1]{\subsection{#1}}
\newcommand{\Acknow}[1]{\par\vspace{5mm}{\bf Acknowledgements.} #1}
\pagestyle{myheadings}

\def\thefootnote{\fnsymbol{footnote}}
\begin{center}

{\Large\bf The Optical Gravitational Lensing Experiment.\\
Variable Stars in Star Clusters of the Magellanic Clouds.\\ 
I.~Eclipsing Systems in the Clusters of the SMC\footnote{Based on 
observations obtained with the 1.3~m Warsaw telescope at the Las
Campanas  Observatory of the Carnegie Institution of Washington.}}

\vskip 1cm

{\bf G.~~P~i~e~t~r~z~y~\'n~s~k~i,~~ and~~A.~~U~d~a~l~s~k~i}
\vskip5mm
{Warsaw University Observatory, Al.~Ujazdowskie~4, 00-478~Warszawa, Poland\\
e-mail: (pietrzyn,udalski)@sirius.astrouw.edu.pl}

\end{center}

\Abstract{The list of 127 eclipsing stars in optical coincidence with star 
clusters from the SMC is presented. It was prepared using the catalogs of 
eclipsing systems and star clusters from the SMC based on observations 
collected during the OGLE-II microlensing project.

Location of 12 eclipsing stars in the color-magnitude diagram of clusters 
allows to exclude their membership. Photometric data of 73 systems support 
their membership. The remaining 42 objects were found in loose, faint 
clusters and therefore no conclusive statement about their membership can be 
made. All presented data are available from the OGLE archive.} {~}

\Section{Introduction}
Both binary eclipsing stars and star clusters are known as a very important 
source of information concerning fundamental problems of stellar 
astrophysics. Binary stars provide opportunity to derive basic stellar 
parameters such as dimensions, masses, distances and stellar composition. 
Observations of such stars allow for empirical test of stellar evolution 
models. Comprehensive discussion of these topics can be found in Andersen 
(1991). 

Binary stars located in star clusters are especially well suited for many 
detailed studies of stellar evolution and dynamics. They may provide 
independent information on cluster distance, age, chemical composition, 
interstellar absorption and formation environment. Observations of binary 
systems in clusters help in studies of differential star formation, anomalies 
in the initial composition and the role of these stars in the dynamical 
evolution of star clusters. Moreover, analysis of distribution of binary stars 
in clusters may help in better understanding their formation and evolution. 

An extensive list of binary stars in optical coincidence with galactic open 
clusters was provided by Gimenez and Clausen (1996). Unfortunately, no such 
catalogs exist for clusters from other galaxies, in particular the SMC. Only 
sporadic information about binary stars located in the SMC star clusters can 
be found in the literature (Hodge and  Wright 1975, Sebo and Wood 1994). 

The main goal of this paper is to provide for further detailed studies a list 
of eclipsing systems detected during the OGLE-II microlensing search and 
located in vicinity of star clusters. In the next section the  observational 
material and used catalogs are described. The resulting list of 127 eclipsing 
systems from clusters of the SMC is presented in Section~3. 

\Section{Observations}
All photometric data were obtained with the 1.3~m Warsaw telescope located at 
the Las Campanas Observatory, Chile, which is operated by the Carnegie 
Institution of  Washington. Detailed description of the instrumental system 
and reduction techniques were presented by Udalski, Kubiak and Szyma{\'n}ski 
(1997). Udalski \etal (1998a) published the {\it BVI}-maps of the Small 
Magellanic Cloud. These maps contain precise photometric data for about 2.2 
million stars from the central region of the SMC bar. Using these data the 
catalog of 238 star clusters was constructed by Pietrzy{\'n}ski \etal (1998). 
Udalski \etal (1998b) presented the sample of 1459 eclipsing binary stars from 
the observed region of the SMC. This catalog contains objects brighter than 
20~mag in the {\it I}-band with periods ranging from about 0.3 to 250~days. 
Its average completeness is about 80\%. 

\Section{Eclipsing Stars in the SMC Clusters}
The catalogs of star clusters and eclipsing systems described in the previous 
Section make it possible to select eclipsing binary stars that are located 
close to the star clusters. We searched for eclipsing systems located at the 
distance smaller than 1.5 radius of a given cluster (Pietrzy{\'n}ski \etal 
1998). Altogether 124 objects passed this criterium. This group was extended 
by 3 eclipsing stars from the region of NGC~346. This cluster is situated  at 
the edge of the observed field and it was not included in the catalog  of 
clusters from the SMC (Pietrzy{\'n}ski \etal 1998). However, all three 
eclipsing stars were found to be in the region of the cluster and we decided 
to include them to our list. Resulting list comprises 127 objects. Their 
description and most important data are given in Table~1.  
\renewcommand{\TableFont}{\scriptsize}
\MakeTableSep{|c|r|c|c|c|c|r|r|c|}{10cm}{Eclipsing systems in clusters from
the SMC}{
\hline
Name  &\multicolumn{1}{|c|}{star}& D  & P & JD & $I$ & $B-V$ & $V-I$ & comments \\
OGLE-CL-&\multicolumn{1}{|c|}{number}&$[R_{\rm CL}]$& [days] &[2450000+]& [mag] & [mag] & [mag] & \\ \hline
SMC0003 &    58960 &  0.914 &    5.12291 &  628.65562 &  16.710 &  $-0.063$ &  $-0.007$ & nm \\ 
SMC0013 &    79559 &  1.068 &   41.65820 &  629.73121 &  16.640 &  $ 0.931$ &  $ 1.118$ & nm \\ 
SMC0014 &    84947 &  0.788 &    0.70806 &  626.28625 &  17.942 &  $-0.118$ &  $-0.029$ & -- \\ 
SMC0016 &    28139 &  1.029 &   21.24938 &  638.79046 &  17.044 &  $ 0.199$ &  $ 0.388$ & nm \\ 
SMC0017 &    63551 &  0.622 &    2.96915 &  627.65183 &  16.709 &  $-0.042$ &  $-0.028$ & pm \\ 
SMC0018 &    63551 &  0.050 &    2.96915 &  627.65183 &  16.709 &  $-0.042$ &  $-0.028$ & pm \\ 
SMC0026 &   193411 &  0.609 &   28.41870 &  637.74818 &  17.197 &  $ 0.195$ &  $ 0.555$ & pm \\ 
SMC0026 &   193413 &  0.222 &    0.76458 &  627.10658 &  17.013 &  $-0.001$ &  $ 0.124$ & -- \\ 
SMC0026 &   193416 &  0.115 &    0.90088 &  626.33819 &  16.706 &  $-0.127$ &  $-0.053$ & -- \\ 
SMC0026 &   194000 &  0.525 &    0.65031 &  626.87302 &  18.635 &  $-0.058$ &  $ 0.042$ & -- \\ 
SMC0030 &    26565 &  0.688 &    4.05740 &  637.99373 &  17.917 &  $ 0.103$ &  $ 0.302$ & -- \\ 
SMC0032 &     3419 &  0.901 &    1.35351 &  629.08725 &  18.743 &  $-0.027$ &  $ 0.110$ & pm \\ 
SMC0034 &    75491 &  0.597 &    1.46307 &  623.09817 &  17.866 &  $ 0.050$ &  $ 0.130$ & -- \\ 
SMC0037 &    82098 &  0.757 &   28.96294 &  673.11576 &  17.230 &  $ 0.655$ &  $ 0.918$ & -- \\ 
SMC0038 &   113853 &  0.353 &    1.32070 &  624.88190 &  17.319 &  $-0.035$ &  $ 0.057$ & pm \\ 
SMC0038 &   113885 &  1.274 &    2.38320 &  624.37028 &  17.122 &  $-0.130$ &  $-0.109$ & pm \\ 
SMC0038 &   114287 &  0.800 &    0.60152 &  623.33710 &  18.381 &  $-0.008$ &  $ 0.167$ & pm \\ 
SMC0039 &   103852 &  0.456 &    2.59490 &  627.43922 &  17.548 &  $-0.085$ &  $-0.116$ & pm \\ 
SMC0047 &   182601 &  1.075 &    1.38436 &  625.28909 &  15.585 &  $-0.152$ &  $-0.130$ & -- \\ 
SMC0048 &    17574 &  1.180 &    1.44567 &  468.87371 &  18.521 &  $-0.080$ &  $-0.056$ & -- \\ 
SMC0048 &    21245 &  0.674 &    2.21727 &  632.99960 &  17.190 &  $-0.067$ &  $-0.081$ & -- \\ 
SMC0049 &    11416 &  0.731 &    1.84072 &  469.75719 &  13.735 &  $-0.231$ &  $-0.230$ & pm \\ 
SMC0049 &    11486 &  0.661 &    1.25288 &  468.01034 &  16.009 &  $-0.175$ &  $-0.214$ & pm \\ 
SMC0054 &    95337 &  0.401 &    0.90458 &  467.10020 &  17.050 &  $-0.105$ &  $-0.143$ & pm \\ 
SMC0054 &    95557 &  0.580 &    2.42120 &  469.22869 &  17.412 &  $-0.082$ &  $-0.085$ & pm \\ 
SMC0054 &    95581 &  1.100 &    0.75725 &  466.87057 &  17.948 &  $ 0.073$ &  $ 0.157$ & pm \\ 
SMC0054 &    96013 &  0.995 &    2.75340 &  468.04339 &  18.384 &  $-0.061$ &  $ 0.047$ & pm \\ 
SMC0056 &   145588 &  1.268 &  138.59136 &  634.71506 &  15.730 &  $ 1.056$ &  $ 1.349$ & -- \\ 
SMC0057 &   123484 &  1.046 &    4.09589 &  468.55460 &  17.136 &  $-0.051$ &  $ 0.028$ & -- \\ 
SMC0058 &   140701 &  0.896 &    3.62537 &  471.12030 &  15.215 &  $-0.064$ &  $-0.054$ & nm \\ 
SMC0058 &   141743 &  1.233 &    1.08615 &  466.88737 &  18.645 &  $ 0.128$ &  $ 0.172$ & pm \\ 
SMC0059 &   202826 &  0.425 &    0.84967 &  467.37147 &  18.233 &  $-0.071$ &  $-0.032$ & pm \\ 
SMC0060 &   170658 &  0.881 &    0.61595 &  467.76091 &  17.357 &  $-0.117$ &  $-0.102$ & pm \\ 
SMC0061 &   185385 &  0.636 &    2.58917 &  468.60764 &  17.626 &  $-0.076$ &  $-0.047$ & pm \\ 
SMC0063 &   202153 &  0.874 &    4.60677 &  468.15759 &  14.268 &  $-0.215$ &  $-0.150$ & -- \\ 
SMC0064 &   208049 &  0.986 &    3.02987 &  471.97126 &  16.094 &  $-0.122$ &  $-0.102$ & pm \\ 
SMC0066 &   266513 &  0.813 &    1.10752 &  467.15386 &  17.985 &  $-0.028$ &  $ 0.069$ & pm \\ 
SMC0067 &   316708 &  0.820 &    7.11698 &  477.77823 &  14.337 &  $-0.162$ &  $ 0.014$ & nm \\ 
SMC0068 &   180064 &  0.634 &    2.51534 &  471.56185 &  15.952 &  $-0.153$ &  $-0.141$ & pm \\ 
SMC0068 &   180197 &  0.830 &    1.39565 &  468.93846 &  17.147 &  $-0.126$ &  $-0.090$ & pm \\ 
SMC0068 &   180519 &  0.796 &    1.45855 &  468.27431 &  18.111 &  $-0.026$ &  $ 0.007$ & pm \\ 
SMC0068 &   181045 &  0.563 &    2.17249 &  467.96586 &  18.945 &  $-0.077$ &  $ 0.078$ & pm \\ 
SMC0068 &   181436 &  1.274 &    1.09748 &  467.30007 &  18.532 &  $ 0.105$ &  $ 0.197$ & pm \\ 
SMC0068 &   260939 &  1.163 &    2.30191 &  468.37916 &  17.157 &  $-0.074$ &  $ 0.007$ & pm \\ 
SMC0068 &   261704 &  1.100 &    0.51489 &  467.07221 &  18.253 &  $-0.003$ &  $ 0.065$ & pm \\ 
SMC0068 &   262198 &  1.008 &    0.93415 &  468.23077 &  18.670 &  $-0.088$ &  $ 0.062$ & pm \\ 
SMC0068 &   263233 &  0.981 &    0.59216 &  467.23901 &  19.182 &  $ 0.025$ &  $ 0.156$ & pm \\ 
SMC0069 &   271332 &  0.874 &    2.17910 &  469.20830 &  17.093 &  $-0.074$ &  $-0.016$ & pm \\ 
SMC0069 &   271839 &  1.167 &    1.18524 &  466.80629 &  17.825 &  $-0.074$ &  $ 0.050$ & pm \\ 
SMC0072 &    17316 &  0.624 &    2.32397 &  469.35747 &  15.216 &  $-0.021$ &  $ 0.077$ & pm \\ 
SMC0072 &    17965 &  0.116 &    3.35528 &  468.69439 &  17.604 &  $-0.071$ &  $-0.011$ & pm \\ 
SMC0074 &    50745 &  1.178 &    0.55453 &  467.32023 &  18.657 &  $-0.120$ &  $ 0.007$ & pm \\ 
\hline
}

\setcounter{table}{0} 
\MakeTableSep{|c|r|c|c|c|c|r|r|c|}{10cm}{Continued}{
\hline
Name  &\multicolumn{1}{|c|}{star}& D  & P & JD & $I$ & $B-V$ & $V-I$ & comments \\
OGLE-CL-&\multicolumn{1}{|c|}{number}&$[R_{\rm CL}]$& [days] &[2450000+]& [mag] & [mag] & [mag] & \\ \hline
SMC0075 &    30725 &  0.239 &    1.47601 &  469.19621 &  18.365 &  $ 0.014$ &  $ 0.116$ & pm \\ 
SMC0078 &    42389 &  0.855 &    3.59519 &  469.94571 &  16.990 &  $-0.092$ &  $-0.070$ & pm \\ 
SMC0078 &   131094 &  0.603 &    0.62343 &  466.71493 &  18.719 &  $-0.100$ &  $-0.143$ & pm \\ 
SMC0079 &    94796 &  1.019 &    0.95196 &  468.27280 &  17.925 &  $-0.031$ &  $ 0.007$ & pm \\ 
SMC0081 &   158178 &  1.089 &    2.16926 &  469.28297 &  16.538 &  $-0.201$ &  $-0.199$ & pm \\ 
SMC0081 &   158653 &  1.041 &    0.83367 &  468.00147 &  17.336 &  $-0.086$ &  $-0.079$ & pm \\ 
SMC0082 &   136065 &  0.752 &    1.00718 &  467.27253 &  18.247 &  $-0.015$ &  $-0.012$ & pm \\ 
SMC0085 &   154456 &  0.590 &    0.64416 &  467.61296 &  18.935 &  $-0.017$ &  $ 0.045$ & -- \\ 
SMC0088 &   141994 &  0.983 &    2.30333 &  468.73850 &  17.481 &  $-0.123$ &  $ 0.041$ & -- \\ 
SMC0088 &   142123 &  0.847 &    1.08822 &  466.77418 &  18.122 &  $-0.077$ &  $ 0.000$ & -- \\ 
SMC0088 &   142746 &  0.711 &    1.67802 &  469.17787 &  18.756 &  $-0.071$ &  $-0.042$ & pm \\ 
SMC0089 &   163107 &  0.946 &    0.58075 &  467.57887 &  16.956 &  $-0.119$ &  $-0.065$ & pm \\ 
SMC0089 &   167473 &  0.707 &    6.35793 &  478.55778 &  15.065 &  $-0.220$ &  $-0.140$ & pm \\ 
SMC0089 &   242498 &  0.980 &    3.64020 &  470.31809 &  17.675 &  $-0.150$ &  $-0.090$ & pm \\ 
SMC0089 &   246749 &  0.728 &    1.64432 &  468.67642 &  17.219 &  $-0.085$ &  $-0.002$ & pm \\ 
SMC0090 &   180283 &  0.963 &    0.88876 &  466.90513 &  16.829 &  $-0.209$ &  $-0.087$ & nm \\ 
SMC0091 &   167623 &  1.168 &    1.04318 &  467.04776 &  16.654 &  $-0.202$ &  $-0.102$ & -- \\ 
SMC0092 &   232226 &  1.002 &    1.18793 &  468.04780 &  15.276 &  $-0.264$ &  $-0.165$ & pm \\ 
SMC0092 &   232282 &  0.706 &    2.55614 &  467.69596 &  16.219 &  $-0.127$ &  $-0.096$ & pm \\ 
SMC0094 &   242378 &  0.926 &    0.97398 &  466.82686 &  17.442 &  $-0.012$ &  $ 0.021$ & -- \\ 
SMC0095 &   180306 &  0.471 &    4.59712 &  471.82987 &  17.099 &  $-0.087$ &  $-0.042$ & -- \\ 
SMC0096 &   242378 &  0.617 &    0.97398 &  466.82686 &  17.442 &  $-0.012$ &  $ 0.021$ & -- \\ 
SMC0096 &   242404 &  0.521 &    3.32876 &  467.89294 &  17.957 &  $ 0.031$ &  $ 0.172$ & -- \\ 
SMC0097 &   296902 &  0.577 &    4.04468 &  469.65193 &  16.912 &  $-0.164$ &  $-0.097$ & -- \\ 
SMC0098 &     9109 &  0.497 &    2.47810 &  631.58654 &  18.192 &  $-0.144$ &  $-0.116$ & pm \\ 
SMC0099 &    70933 &  0.484 &    2.05886 &  627.48603 &  17.440 &  $-0.150$ &  $-0.082$ & pm \\ 
SMC0099 &    70971 &  0.899 &    2.24680 &  629.60491 &  17.341 &  $ 0.122$ &  $-0.123$ & pm \\ 
SMC0102 &    52810 &  0.592 &    0.61413 &  627.53633 &  17.648 &  $-0.098$ &  $-0.100$ & nm \\ 
SMC0103 &    91646 &  0.988 &    1.42483 &  628.89832 &  16.801 &  $-0.136$ &  $-0.102$ & -- \\ 
SMC0103 &    91654 &  0.777 &    3.07494 &  628.24188 &  17.025 &  $-0.088$ &  $-0.016$ & -- \\ 
SMC0104 &   110723 &  0.461 &    4.71272 &  632.41390 &  17.802 &  $ 0.060$ &  $ 0.228$ & -- \\ 
SMC0104 &   115177 &  0.928 &    2.14025 &  628.77436 &  16.434 &  $-0.052$ &  $-0.145$ & -- \\ 
SMC0105 &   110129 &  0.385 &    2.35207 &  629.19120 &  16.249 &  $-0.116$ &  $-0.082$ & pm \\ 
SMC0107 &   206212 &  0.479 &    5.58638 &  628.96334 &  16.821 &  $-0.150$ &  $-0.151$ & pm \\ 
SMC0107 &   206520 &  0.298 &    0.93333 &  628.16198 &  18.224 &  $-0.152$ &  $-0.064$ & pm \\ 
SMC0107 &   207302 &  0.563 &    8.43208 &  638.90384 &  18.922 &  $ 0.607$ &  $ 0.915$ & pm \\ 
SMC0113 &    87145 &  0.779 &    2.06187 &  627.85368 &  15.677 &  $-0.116$ &  $-0.055$ & -- \\ 
SMC0117 &   139441 &  0.974 &    1.48244 &  626.98531 &  16.366 &  $-0.118$ &  $-0.079$ & nm \\ 
SMC0123 &    39292 &  1.048 &    0.80302 &  657.71342 &  17.626 &  $-0.185$ &  $-0.066$ & -- \\ 
SMC0126 &    13423 &  0.706 &    1.63250 &  628.75723 &  16.851 &  $-0.132$ &  $-0.103$ & pm \\ 
SMC0126 &    14011 &  0.400 &    3.16417 &  630.94891 &  18.378 &  $ 0.017$ &  $ 0.081$ & pm \\ 
SMC0127 &    42184 &  1.089 &    1.08486 &  628.58889 &  18.365 &  $-0.172$ &  $-0.094$ & -- \\ 
SMC0127 &    87367 &  0.977 &    1.04951 &  628.58483 &  18.816 &  $-0.156$ &  $-0.045$ & -- \\ 
SMC0128 &    79005 &  0.961 &    0.99627 &  628.05609 &  17.888 &  $-0.078$ &  $-0.069$ & pm \\ 
SMC0130 &    89448 &  0.754 &    5.74206 &  637.20271 &  17.238 &  $-0.192$ &  $-0.176$ & -- \\ 
SMC0134 &   169330 &  0.938 &    0.53735 &  628.04710 &  18.876 &  $-0.065$ &  $ 0.018$ & pm \\ 
SMC0135 &   147421 &  0.910 &    0.45477 &  628.52801 &  18.887 &  $-0.077$ &  $ 0.132$ & -- \\ 
SMC0136 &    19778 &  0.328 &    0.90602 &  657.09768 &  17.841 &  $ 9.999$ &  $-0.097$ & -- \\ 
SMC0141 &    52356 &  0.847 &    1.14518 &  629.35668 &  18.032 &  $-0.075$ &  $-0.025$ & pm \\ 
SMC0142 &    70555 &  0.781 &    1.51265 &  630.12421 &  17.746 &  $-0.147$ &  $-0.062$ & pm \\ 
SMC0144 &    31290 &  1.226 &    1.51766 &  628.64208 &  16.466 &  $-0.177$ &  $-0.144$ & pm \\ 
SMC0145 &   110466 &  0.543 &    0.67506 &  628.31409 &  16.624 &  $-0.105$ &  $-0.059$ & pm \\ 
\hline
}
 
\setcounter{table}{0} 
\MakeTable{|c|r|c|c|c|c|r|r|c|}{10cm}{Concluded}{
\hline
Name  &\multicolumn{1}{|c|}{star}& D  & P & JD & $I$ & $B-V$ & $V-I$ & comments \\
OGLE-CL-&\multicolumn{1}{|c|}{number}&$[R_{\rm CL}]$& [days] &[2450000+]& [mag] & [mag] & [mag] & \\ \hline
SMC0145 &   110567 &  0.625 &    1.16732 &  629.60675 &  17.398 &  $-0.009$ &  $ 0.027$ & pm \\ 
SMC0145 &   110574 &  1.085 &    0.69045 &  628.64620 &  18.143 &  $-0.108$ &  $-0.068$ & pm \\ 
SMC0147 &   110466 &  0.391 &    0.67506 &  628.31409 &  16.624 &  $-0.105$ &  $-0.059$ & pm \\ 
SMC0156 &    42477 &  0.876 &    1.14013 &  625.95929 &  17.625 &  $-0.124$ &  $-0.019$ & pm \\ 
SMC0158 &    58186 &  0.577 &   10.51155 &  640.20423 &  18.350 &  $ 0.089$ &  $ 0.148$ & nm \\ 
SMC0158 &    89239 &  0.664 &    5.16689 &  626.61741 &  17.166 &  $-0.095$ &  $-0.054$ & nm \\ 
SMC0158 &    89956 &  0.682 &    1.30050 &  626.09550 &  18.290 &  $-0.042$ &  $ 0.021$ & nm \\ 
SMC0159 &    38611 &  1.075 &    2.16874 &  629.61251 &  17.698 &  $-0.089$ &  $ 0.004$ & nm \\ 
SMC0159 &    68552 &  0.826 &    0.90927 &  627.53293 &  18.482 &  $-0.104$ &  $ 0.033$ & nm \\ 
SMC0161 &    95492 &  1.091 &    3.59535 &  632.62143 &  17.935 &  $ 0.036$ &  $ 0.016$ & -- \\ 
SMC0194 &    16658 &  0.965 &    1.24619 &  468.55853 &  17.414 &  $-0.148$ &  $-0.100$ & pm \\ 
SMC0195 &    27518 &  0.824 &    0.57604 &  467.02297 &  18.563 &  $-0.006$ &  $-0.005$ & pm \\ 
SMC0195 &   106240 &  0.300 &    0.73385 &  467.41721 &  18.104 &  $ 0.047$ &  $ 0.051$ & pm \\ 
SMC0201 &   171289 &  0.809 &   16.84515 &  470.37832 &  18.197 &  $ 0.500$ &  $ 0.797$ & pm \\ 
SMC0206 &    24274 &  0.867 &    1.94059 &  469.03518 &  18.184 &  $-0.019$ &  $ 0.065$ & -- \\ 
SMC0209 &   154048 &  0.355 &   11.52736 &  468.41539 &  18.769 &  $ 0.246$ &  $ 0.430$ & -- \\ 
SMC0216 &   222388 &  1.119 &    6.61099 &  468.62792 &  18.627 &  $ 0.027$ &  $ 0.162$ & -- \\ 
SMC0216 &   297080 &  0.219 &    8.82380 &  469.72748 &  17.536 &  $-0.057$ &  $ 0.048$ & -- \\ 
SMC0226 &    38858 &  0.524 &    1.55266 &  629.72831 &  16.335 &  $-0.093$ &  $-0.089$ & -- \\ 
SMC0234 &   163639 &  0.393 &    0.68523 &  629.17323 &  16.545 &  $-0.184$ &  $-0.124$ & pm \\ 
NGC 346 &   160677 &  --    &   86.42187 &  658.097   & 14.262  &  $ 0.054$ &  $ 0.154$ & -- \\
NGC 346 &   160725 &  --    &    3.10487 &  635.440   & 15.857  &     --    &  $-0.110$ & -- \\
NGC 346 &   160975 &  --    &    0.65143 &  629.968   & 17.927  &  $-0.057$ &  $-0.019$ & --\\
\hline
}

First column of Table~1 contains cluster name. Star ID number and distance 
from the cluster center, $D$, measured in units of cluster radius are 
presented in columns 2 and 3. Period in days, zero epoch corresponding to the 
deeper eclipse, $I$-band magnitude and ${B-V}$ and ${V-I}$ colors extracted 
from the catalog of eclipsing systems (Udalski \etal 1998b) are listed in 
columns 4, 5, 6, 7 and 8. 

\begin{figure}[p]
\vspace*{-4mm}
\psfig{figure=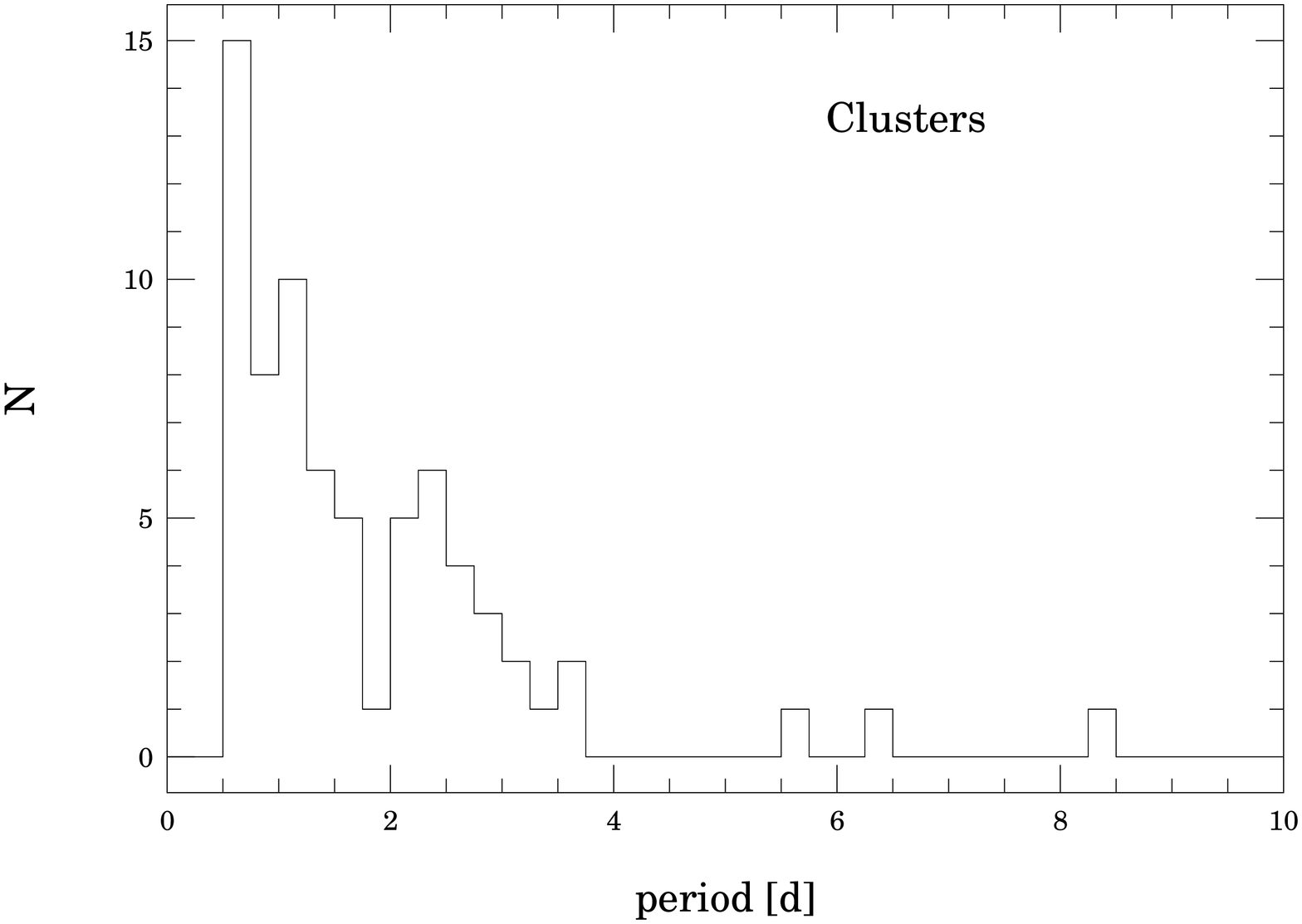,width=13cm,angle=90,clip=}
\vskip-3mm
\FigCap{Distribution of periods of 73 eclipsing stars regarded as probable 
members of the SMC clusters.} 
\psfig{figure=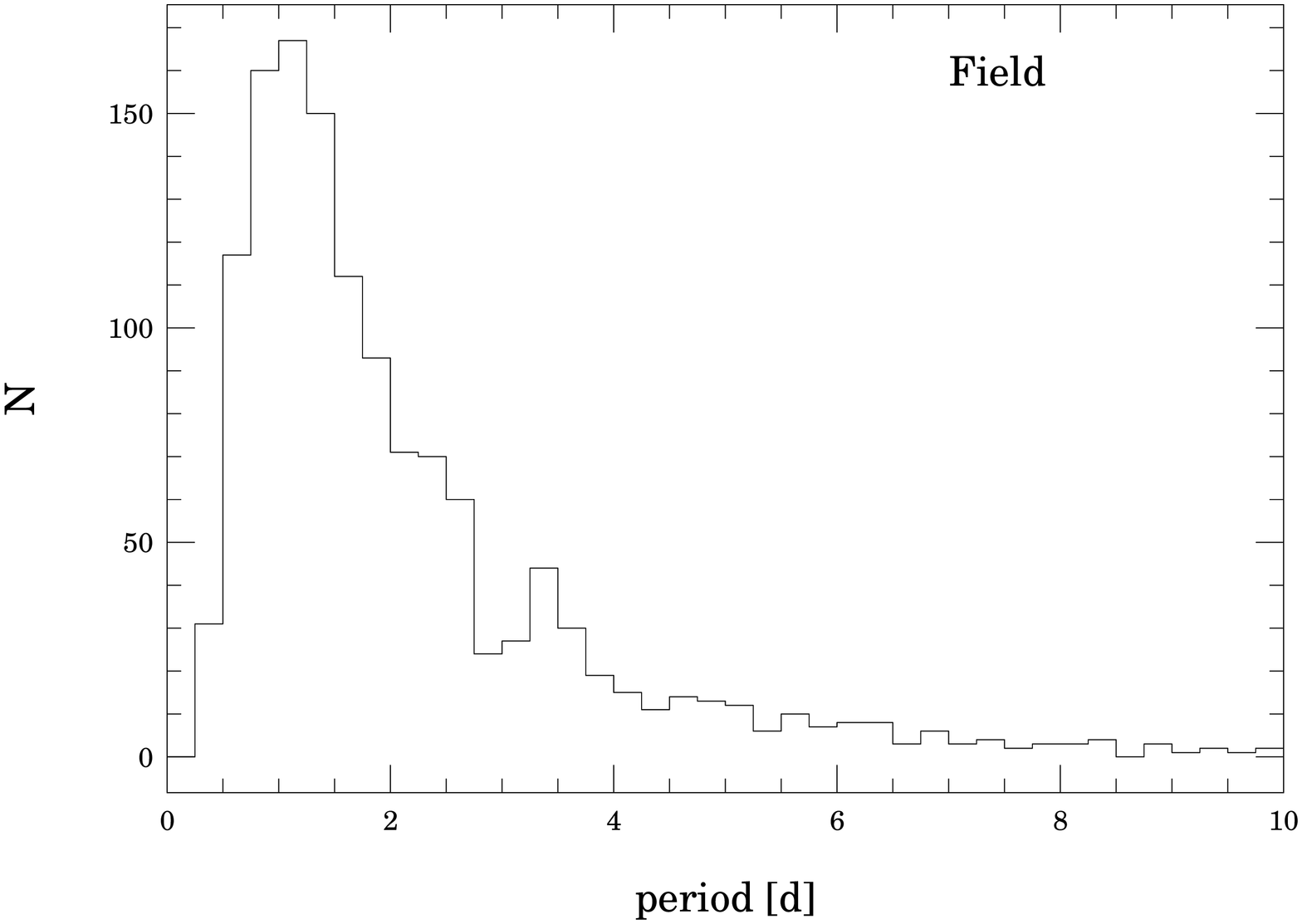,width=13cm,angle=90,clip=}
\vskip-3mm
\FigCap{Distribution of periods of 1332 eclipsing stars from the field of the 
SMC.} 
\end{figure}
Based on location of each eclipsing star in the color-magnitude diagram (CMD) 
of the appropriate cluster preliminary classification was made. 85 eclipsing 
variable stars are located in the vicinity of relatively populous clusters. 
Position of 12 of them in the CMD allows to exclude their membership. They 
were classified as non-members (nm). Location of 73 systems in the CMDs of 
these clusters is consistent with their membership. These stars were designed 
as probable members (pm). The remaining 42 eclipsing systems are found around  
poorly populated clusters and no conclusive statements about their membership 
can be made. Information about cluster membership of eclipsing systems is 
given in column 9 of Table~1. 

Figs.~1 and 2 present distribution of periods of eclipsing binary stars from 
clusters (those classified as probable members) and those from the field of 
the SMC. Similar shape of these distributions suggests common processes of 
formation and disruption of eclipsing  stars located  in the field and 
clusters of the SMC. 

\Section{Summary}
OGLE-II catalogs of eclipsing systems and star clusters from the regions 
covering about 2.4 square degree of the central parts of the SMC were used for 
selection of eclipsing systems located in star clusters. 127 eclipsing 
stars were found  within 1.5 cluster radii. Based on their position in the 
cluster CMDs the preliminary classification was made. 73 and 12 eclipsing 
systems were  labeled as probable members and non-members, respectively. Our 
photometric data do not allow to obtain information on membership of the 
remaining 42 objects. Distribution of periods of eclipsing stars from star 
clusters and field objects of the SMC are similar which may suggest similar 
processes of formation and disruption of eclipsing systems in the field and 
star cluster regions. 

Table~1 and all photometric data can be obtained from the OGLE archive: {\it 
http://www.astrouw.edu.pl/\~{}ftp/ogle}. 

\Acknow{The paper was partly supported by the Polish KBN grant 2P03D00814 to 
A.\ Udalski.  Partial support for the OGLE project was provided with the NSF 
grant  AST-9530478 to B.~Paczy\'nski.}

\end{document}